\documentclass[draft]{agujournal2019}
\usepackage{url} %this package should fix any errors with URLs in refs.
\usepackage{lineno}
\usepackage[finalnew]{trackchanges} %for better track changes. finalnew option will compile document with changes incorporated.
\usepackage{soul}
\usepackage{natbib}
\draftfalse

\journalname{JGR: Space Physics}

\begin{document}

%% ------------------------------------------------------------------------ %%
%  Title
%
% (A title should be specific, informative, and brief. Use
% abbreviations only if they are defined in the abstract. Titles that
% start with general keywords then specific terms are optimized in
% searches)
%
%% ------------------------------------------------------------------------ %%

\title{Six pieces of evidence against the corotation enforcement theory to explain the main aurora at Jupiter}

%% ------------------------------------------------------------------------ %%
%
%  AUTHORS AND AFFILIATIONS
%
%% ------------------------------------------------------------------------ %%

% Authors are individuals who have significantly contributed to the
% research and preparation of the article. Group authors are allowed, if
% each author in the group is separately identified in an appendix.)

% List authors by first name or initial followed by last name and
% separated by commas. Use \affil{} to number affiliations, and
% \thanks{} for author notes.
% Additional author notes should be indicated with \thanks{} (for
% example, for current addresses).

% Example: \authors{A. B. Author\affil{1}\thanks{Current address, Antartica}, B. C. Author\affil{2,3}, and D. E.
% Author\affil{3,4}\thanks{Also funded by Monsanto.}}

\authors{B. Bonfond \affil{1}}
\authors{Z. Yao \affil{2,1}}
\authors{D. Grodent \affil{1}}

% \affiliation{1}{First Affiliation}
% \affiliation{2}{Second Affiliation}
% \affiliation{3}{Third Affiliation}
% \affiliation{4}{Fourth Affiliation}

\affiliation{1}{LPAP, STAR Institute, Universit\'e de Liège, Belgium}
\affiliation{2}{Key Laboratory of Earth and Planetary Physics, Institute of Geology and Geophysics, Chinese Academy of Sciences, Beijing, China.}
%(repeat as many times as is necessary)

%% Corresponding Author:
% Corresponding author mailing address and e-mail address:

% (include name and email addresses of the corresponding author.  More
% than one corresponding author is allowed in this LaTeX file and for
% publication; but only one corresponding author is allowed in our
% editorial system.)

% Example: \correspondingauthor{First and Last Name}{email@address.edu}

\correspondingauthor{Bertrand Bonfond}{b.bonfond@uliege.be}

%% Keypoints, final entry on title page.

%  List up to three key points (at least one is required)
%  Key Points summarize the main points and conclusions of the article
%  Each must be 100 characters or less with no special characters or punctuation and must be complete sentences

% Example:
% \begin{keypoints}
% \item	List up to three key points (at least one is required)
% \item	Key Points summarize the main points and conclusions of the article
% \item	Each must be 100 characters or less with no special characters or punctuation and must be complete sentences
% \end{keypoints}

\begin{keypoints}
\item The corotation enforcement current system currently is the \change{mainstream}{most widely-accepted} explanation for the main auroral emissions at Jupiter
\item We \change{expose}{present} six observational pieces of evidence that this theory is \add{probably} not the main explanation for these auroral emissions 
\item Improved theories should account for the local time variations in the magnetosphere and the importance of the plasma waves for the creation of auroral emissions.
\end{keypoints}

%% ------------------------------------------------------------------------ %%
%
%  ABSTRACT and PLAIN LANGUAGE SUMMARY
%
% A good Abstract will begin with a short description of the problem
% being addressed, briefly describe the new data or analyses, then
% briefly states the main conclusion(s) and how they are supported and
% uncertainties.

% The Plain Language Summary should be written for a broad audience,
% including journalists and the science-interested public, that will not have 
% a background in your field.
%
% A Plain Language Summary is required in GRL, JGR: Planets, JGR: Biogeosciences,
% JGR: Oceans, G-Cubed, Reviews of Geophysics, and JAMES.
% see http://sharingscience.agu.org/creating-plain-language-summary/)
%
%% ------------------------------------------------------------------------ %%

%% \begin{abstract} starts the second page

\begin{abstract}
The most remarkable feature of the ultraviolet auroras at Jupiter is the ever\add{-}present and almost continuous curtain of bright emissions centered on each magnetic pole and called the main emissions. According to the \change{classical}{widely-accepted} theory, it results from an electric current loop transferring momentum from the Jovian ionosphere to the magnetospheric plasma. However, predictions based on \change{these mainstream models have}{this theory has} been recently challenged by observations from Juno and the Hubble Space Telescope. Here we review the main contradictory observations, expose their implications for the theory and discuss promising paths forward.
\end{abstract}

\section*{Plain Language Summary}
The powerful auroras at Jupiter are very different from those at Earth and the mechanisms generating them differ as well. Their most obvious feature\remove{s} is a relatively continuous auroral curtain surrounding the magnetic poles. The classical explanation for its presence involves an electric current system that \change{allows}{causes} particles in the magnetosphere to rotate with planet. While these models explain some characteristics of the auroras, recent observations from the NASA Juno spacecraft and the Hubble Space Telescope challenge this theoretical framework. 

% These auroras are mainly caused by electrons flowing along magnetic field line and hitting the neutral atmosphere in the polar regions.

%% ------------------------------------------------------------------------ %%
%
%  TEXT
%
%% ------------------------------------------------------------------------ %%

%%% Suggested section heads:
% \section{Introduction}
%
% The main text should start with an introduction. Except for short
% manuscripts (such as comments and replies), the text should be divided
% into sections, each with its own heading.

% Headings should be sentence fragments and do not begin with a
% lowercase letter or number. Examples of good headings are:

% \section{Materials and Methods}
% Here is text on Materials and Methods.
%
% \subsection{A descriptive heading about methods}
% More about Methods.
%
% \section{Data} (Or section title might be a descriptive heading about data)
%
% \section{Results} (Or section title might be a descriptive heading about the
% results)
%
% \section{Conclusions}

\section{Introduction}

\begin{figure}
 \includegraphics[width=\textwidth]{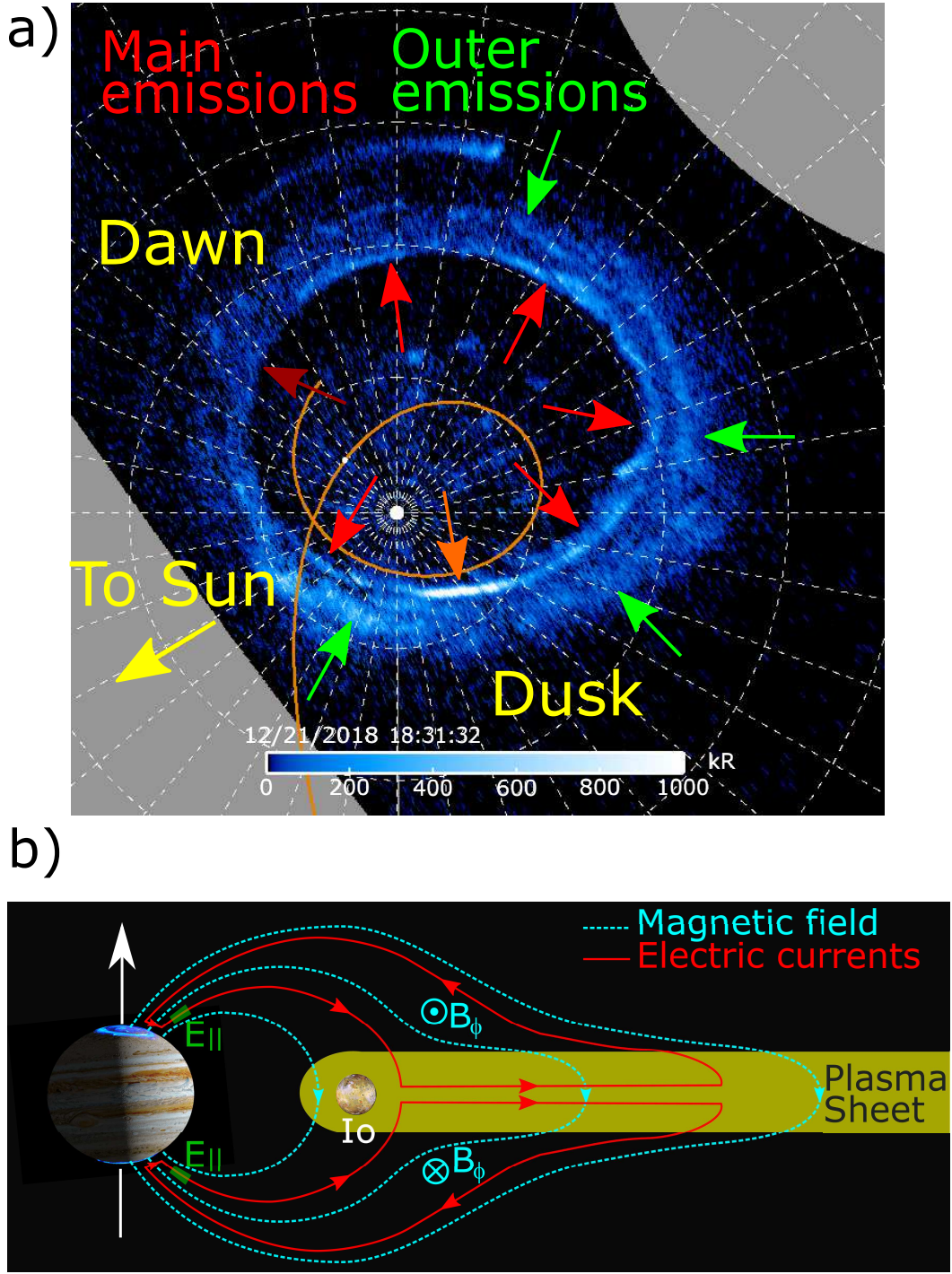}
 \caption{Panel a) A typical polar projection of the southern UV aurora as seen by Juno-UVS on 21 December 2018. The red arrows highlight the main emissions, which are the brightest on the dusk flank (orange) and the dimmest in the pre-noon sector (dark red). The green arrows highlight the outer emissions. The polar emissions were weak on that particular day. Panel b) Classical scheme of the corotation enforcement currents model to explain the main auroral emissions at Jupiter (After \citet{cowley_origin_2001}). The dashed cyan lines represent the magnetic field and the solid red lines represent the electric currents. The electric fields accelerating the electrons into the aurora are shown in green.}
 \label{Figure1}
 \end{figure}

The ultraviolet (UV) auroras at Jupiter can be separated into three almost equally powerful components \citep{nichols_variation_2009, grodent_jupiters_2018}: 1) the main emissions (ME), which \change{are forming}{form} an almost continuous curtain of auroral emissions around the magnetic pole, 2) the polar emissions located poleward of the ME and 3) the equatorward, or outer, emissions, essentially \change{comprised}{confined} between the ME and Io's footpath (Figure \ref{Figure1}a). The auroral footprints of the Galilean moons, are often cited as a fourth component, even if their total emitted power is much smaller ($\sim$30 GW for the Io footprint \citep{bonfond_evolution_2013}, compared to $\sim$500 GW for the ME \citep{grodent_jupiters_2018}). The ME magnetically maps to distances typically ranging from 20 to 60 Jovian radii ($R_J$) \citep{vogt_improved_2011}, though in rare instances, \change{this distance}{the lower boundary} dropped to near to the orbit of Ganymede (15$R_J$) \citep{bonfond_auroral_2012}. Since it became clear that the main emissions \change{did correspond}{corresponded} neither to the open-closed field line boundary (or the outer-most magnetosphere) nor to the Io torus \citep{dols_ultraviolet_1992}, the most widely accepted explanation for this auroral feature involves a large scale current system coupling the magnetospheric plasma to the ionosphere \citep{hill_inertial_1979,hill_jovian_2001, cowley_origin_2001,southwood_new_2001}. 

According to this theoretical frame, the current system transfers momentum from the ionosphere to the plasma sheet. It flows radially outward in the plasma sheet and the $J \times B$ force accelerates the magnetospheric plasma towards corotation with the planet. At the other end of the circuit, equatorward\add{-flowing} Pedersen currents slow down the charged particles in the ionosphere, which interact with the rest of the upper atmosphere via ion-neutral collisions. Field\add{-}aligned currents flow between these two sections of the loop, upward from the ionosphere to the magnetosphere in the middle magnetosphere and in the opposite direction in the outer magnetosphere (Figure \ref{Figure1}b). In the plasma sheet, this current system starts at Io's orbit, where fresh plasma is injected in the magnetosphere from the volcanic moon's neighbourhood. This plasma then progressively migrates outward to be eventually released in the Jovian magnetotail. As the radial distance increases\change{ and}{,} in the absence of additional forces\remove{,} the conservation of the angular momentum dictates that the angular velocity of the plasma would decrease. Thus, to maintain corotation, the required momentum transfer from the ionosphere to the magnetosphere increases, as do the currents. Models predict the field\add{-}aligned currents peak at a distance close to the region where the system becomes unable to maintain full corotation with the planet, also known as the corotation breakdown distance. In the region where the upward currents peak \add{and based on the} \citep{knight_parallel_1973} \add{kinetic theory or a modified version of it }\citep{ray_current-voltage_2009}, field\add{-}aligned potential\add{s} are expected to form and accelerate electrons into the atmosphere, causing the main auroral emissions. \add{It should be noted that, while it is not a formal requirement of the corotation enforcement theory, most models  (i.e. individual published studies based on this theoretical framework) are axisymmetric, or at least locally axisymmetric, meaning that they assume that net azimuthal fluxes can be neglected.}

Many observations gathered either by the spacecraft that have visited the Jovian system through the years or by \change{Earth based}{Earth-based} telescopes appear to support some \change{elemental}{elementary} processes in this framework. First, the magnetospheric plasma at Jupiter is either in full corotation with the planet, or, at least, significantly rotating with it, indicative that momentum is indeed transferred from the ionosphere/thermosphere to the magnetosphere. It should however be noted that this is also true for Saturn, and yet, the associated current system does not give rise to significant auroras (the auroras at Saturn are mainly caused by other processes \citep[see review in][]{stallard_saturns_2014}). Then, sub-corotation and velocity shears in the polar ionosphere of Jupiter have been observed, indicative of a torque being exerted on the inner polar regions of the ionosphere \citep[e.g.][]{johnson_jupiters_2017}. It is also noteworthy that these models predict a \change{location}{range of magnetic latitudes} of the auroral emissions and a typical brightness consistent with the observations. The idea that corotation enforcement currents drive the ME also provides an explanation for the usually dimmer main emissions on their pre-noon section, named the discontinuity \citep{radioti_discontinuity_2008,ray_local_2014}. As the shape of the dayside magnetopause forces the plasma \add{to flow radially inward} in the dawn-side magnetosphere, its azimuthal velocity increases and the need for momentum transfer decreases. The field\add{-}aligned currents inferred from Galileo magnetic field measurements in the equatorial plane are also minimum in this sector \citep{khurana_influence_2001}. Furthermore, the equatorward expansion of the main emissions during \add{a} time interval during which the mass outflow rate is \change{expected}{believed} to have increased is also consistent with the theory \citep{bonfond_auroral_2012}. Finally, the observation of the relationship between the precipitating energy flux in the main emissions and the mean electron energy were found to be consistent with the Knight-like relationship expected for quasi-static electric fields \citep{gustin_energy-flux_2004,gerard_color_2016}\change{, even if}{. It should however be noted that } \citet{clark_precipitating_2018} showed that Alfv\'enic acceleration could lead to the same kind of relationship.

However, while this \change{model}{theory} is widely accepted \change{to be}{as} the explanation for the main auroral emissions and despite its successes, several observations contradicting the predictions of this theoretical framework concerning the main auroral emissions have \remove{recently} started to accumulate. Some of these observations \change{were actually}{have been} known for a long time, while others were recently revealed by the NASA Juno mission.

%\add{The first two pieces of evidence presented here (sections }\ref{Dd_particles} \add{and} \ref{FA_currents} \add{) challenge the relevance of the assumed longitudinal (local) axisymmetry and North-South symmetry of the models of the current system giving rise to these auroral emissions. While these assumptions are not strict requirements of the corotation enforcement theory, their implementations in the vast majority of the CE models appears less benign than expected.} \add{The next three section } \add{Section }\ref{Dd_aurora} \add{discusses how the azimuthal asymmetries affect the main auroral emissions. The next two sections  }

\add{Sub-sections 2.1 and 2.2 discuss the limits of the local axisymmetry and North-South symmetry hypothesis generally adopted in the models of the current system giving rise to these auroral emissions. The next three sub-sections discuss observations challenging the link between radial currents and auroral intensity. And the final section discusses the relevance of quasi-static potentials to generate the ME.}

%Text here ===>>>

\section{Six pieces of evidence against the "corotation enforcement" explanation for the main emissions}

\subsection{Dawn/dusk asymmetries in the particle velocity \note{previously Section 2.4}}
\label{Dd_particles}
In \change{axisymmetric models}{models assuming independent longitudinal cuts through the magnetosphere (local axisymmetry)}, the velocity of the particles and the azimuthal component of the magnetic field (i.e. the bend\add{-}back) are expected to be anti-correlated. However, comparisons of the dawn and dusk flanks of the Jovian magnetosphere show that the magnetic field bend\add{-}back is larger in the dawn flank \citep{khurana_global_2005} as well as the velocity of the charged particles \citep{krupp_global_2001}. When considering the three dimensional shape of the magnetosphere, this actually make\add{s} much sense. As the field lines are increasingly stretched in the magnetotail, the plasma's angular velocity decreases. Then, on the dawn side, the field lines are still considerably stretched backward (compared to the dusk side), but the particles angular velocity increases as the particles are now moving radially inward. \remove{This illustrates again the limitations of axisymmetric models with regard to local time effects.} 

\subsection{Field\add{-}aligned currents are fragmented and asymmetric \note{previously Section 2.5}}
\label{FA_currents}
The first Juno observations of the magnetic field above the Jovian auroras did not reveal the strong field\add{-}aligned currents expected from the theory \citep{connerney_juno_2017}, but a later analysis covering the first 11 Juno perijoves did reveal significant currents, with a combined mean value of 82 MA for the two hemispheres, which is in line with the estimates of the radial currents in the magnetosphere \citep{kotsiaros_birkeland_2019}. However, they found that the current did not take the form of thin and regular current shells, but were fragmented and confined \change{on}{in} longitudinal extent. \add{While this fragmentation is generally overlooked in the theoretic models, it appear, at least qualitatively consistent with the strong plasma turbulence observed in the equatorial plane by Galileo} \citep{mauk_equatorial_2007}. Another \change{unexpected}{striking} feature was the strong asymmetry between the two hemispheres, with southern currents being \change{approxymately}{approximately} twice as large as in the north (58 MA compared to 24 MA). They attributed this difference to the magnetic field asymmetries leading to differences in the Pedersen conductivity between the two polar ionospheres. \add{However, conjugate observations of the ME by HST did not lead to a clear cut relationship between the surface magnetic field and the emitted power in the UV}\citep{gerard_hubble_2013}\add{. The large asymmetries of the Jovian magnetic field have been neglected in theoretical models, including  magneto-hydrodynamic (MHD) simulations,  so far but they appear to have important consequences on both the current and the aurora.}

\subsection{Dawn/dusk brightness asymmetry \note{previously Section 2.3}}
\label{Dd_aurora}
\change{One of the}{The} most direct \add{pieces of} evidence of the radially outward flowing currents in the plasma sheet is the azimuthal bend\add{-}back of the magnetic field. This angle is larger on the dawn flank of the magnetosphere than on the dusk flank. As a consequence, the main emissions are also expected to be brighter \change{on}{at} dawn \citep{ray_local_2014}. However, a comparison of the dawn and dusk sides of the main emissions based on Hubble Space Telescope observations showed that the dusk side is typically 3 times brighter than the dawn side \citep{bonfond_far-ultraviolet_2015}. A possible explanation is that, in addition to the corotation enforcement currents, another current system of the same magnitude and linked to the partial ring current in the magnetotail also feeds into the auroral regions. It would consistently strengthen the total net field\add{-}aligned currents on the dusk side and weaken the currents on the dawn side. Analysing the equatorial magnetic field measurements of the whole Galileo mission, \citet{lorch_local_2020} also concluded that azimuthal currents play a key role in determining the location of the field\add{-}aligned currents.
Furthermore, \citet{vogt_solar_2019} noted that the dawn-dusk discrepancy on the bend\add{-}back angle is even larger during solar wind compressions, which should lead to a brightening of the dawn arc of the ME but a dimming of the dusk arc if the corotation enforcement current were driving the main auroral emissions. Again, this is contrary to the observations, as the ME brighten at all local times during compressions of the magnetopause \citep{yao_auroral_2020}.

\subsection{Global auroral brightening with solar wind compression \note{previously Section 2.1}}
\label{Compressions}
One of the \remove{first} prediction\add{s} of the corotation enforcement currents models concerned the response to solar wind compressions and expansions. All these models predict that the ME aurora would dim as a response to a solar wind compression\add{, due to the smaller size of the magnetosphere, to the increased angular velocity of the plasma which is pushed inward and the lower resulting currents} \citep{southwood_new_2001,cowley_modulation_2003, cowley_origin_2001}. The first versions only considered steady state systems. A later iteration took the time variations into consideration\remove{s} \citep{cowley_modulation_2007}. 

Observations of the infrared $H_3^+$ aurora before the Ulysses Jupiter fly-by showed an increase of the total emitted power with \remove{the} increase of \remove{the} solar wind \add{ram pressure} \citep{baron_solar_1996}. It was\add{,} however\add{,} not clear at the time that this increase was due to \add{an intensification of }the main emissions, or whether it was related to a brightening of other regions. Studies based on Hisaki observations of the total auroral power in the ultraviolet reached the same conclusion \citep{kita_characteristics_2016}. In other wavelengths (e.g. \citet{zarka_low-frequency_1983,gurnett_control_2002} for the radio hectometric emissions, \citet{dunn_impact_2016} for the X-rays owing to ion precipitation and \citet{sinclair_brightening_2019} for the infrared hydrocarbon emissions), increase of the auroral activity have also been found to correlate with compressed solar wind conditions. It should be noted that these indices possibly involve processes taking place poleward of the ME, which may or may not be correlated with the ME. Analysis of the response of the UV aurora to a solar wind compression prior to the Cassini Jupiter fly-by showed that the main emissions brightened during a solar wind compression \citep{nichols_response_2007}. However, the exact timing of the response remained unclear, as the model of \citet{cowley_modulation_2007} predicted a possible brief ME enhancement right after the arrival of a compressed solar wind, before a prolonged dimming of the auroral emissions. \change{Later}{Subsequent} HST observations of the aurora during \change{either}{both} the New Horizons fly-by  \citep{nichols_observations_2009,clarke_response_2009} \change{or}{and} the arrival of Juno \cite{nichols_response_2017} suggested that some auroral brightenings are consistent with intervals of solar wind compressions. A recent study including observations from both Hisaki and the UV spectrograph on board Juno also described brightenings correlated with the solar wind compressions, but concluded that the exact timing of the brightening lagged the arrival of large solar wind shocks \citep{kita_jovian_2019}. They also found that the amplitude of the brightening did not scale with the disturbance of the dynamic pressure. In summary, these studies either conclude that the ME brightens with the arrival of a compression region, or conclude that the timing of the response is unclear, but none of them report the dimming expected from the theory. 

\citet{yao_auroral_2020} observed the aurora with the Hubble Space Telescope as Juno was on the dawn flank of the magnetosphere. Juno encountered \remove{several time} the magnetopause \add{several times} during \remove{time} intervals of compressed magnetosphere. Each time, the main emissions significantly brightened at all local times. Unlike all previous studies, this one does not rely on any propagation model of the solar wind, but directly assess\add{es} the state of the magnetosphere. It is also remarkable that even the noon sector, which is where the compression effects should be the clearest, brightened compared to the quiet case. This study also confirms that hectometric radio emissions are systemically enhanced during solar wind compression. Finally, it should be noted that, while non-resolved enhancements of the auroras do not guarantee that the ME is the auroral component that caused it \citep[see counterexample in][associated with internally driven reconfigurations]{kimura_transient_2015}, the \add{spatially resolved} enhancements of the ME seen by HST \add{(}and Juno-UVS\add{)} result in an enhancement of the total power compatible with those Hisaki and others observed simultaneously to solar wind shocks. 

\add{Two recent MHD simulations of the response of the Jovian magnetosphere and aurorae showed contrasted results. The} \citet{chane_how_2017} \add{simulation shows a brightening of the ME at all local times, which is compatible with the observations of brightened ME, but contrary to expectations from the CE theory. On the other hand, } \citet{sarkango_global_2019} \add{found that the field-aligned currents decrease on the dayside, in accordance with the CE theory, but incompatible with the observed brightening of the ME from HST. It should be noted that MHD simulations are complex tools, modeling more processes that just the corotation enforcement currents and it is often difficult to disentangle which elementary process is causing which specific feature in the model output. Moreover, the fact that the two simulations provide contradictory results indicates that the chosen set of hypotheses, numerical schemes and boundary conditions might considerably affect the end result.}

%\subsection{Noon brightening with solar wind compression}

\subsection{Brightness variations as a response to magnetic loading/unloading \note{previously Section 2.2} }
\label{Loading_unloading}

\citet{yao_relation_2019} directly compared the azimuthal and radial stretching of the dawn-side magnetic field as measured by Juno to the auroral output. During a time interval for which the magnetosphere was compressed, they noted that the auroras and the ME in particular were brighter than during quiet times. \add{It should be noted that the auroral morphology was distinct from dawn storms, probably associated with large scale tail reconnection} \citep{yao_auroral_2020, bonfond_substorm-like_2020}\add{.} They also noted that the stretching of the magnetic field\remove{,} or\add{,} \remove{said} in other words, the loading of energy in the magnetic field, oscillated during this interval. And, contrary to \change{classical}{widely-accepted} theoretical expectation\add{s}, the aurora and the radio kilometric emissions increased during the \add{magnetic} unloading phases, as if the magnetic energy was converted into particle energy, similarly to what is observed on Earth.

\subsection{Quasi-static potentials are not the main driver for the ME}
\label{Qs_potentials}

One of the main finding\add{s} of the Juno mission so far is the ubiquity of \remove{the} stochastic acceleration processes for the charged particles in the polar regions. At Earth, the most steady and brightest auroral emissions are related to quasi-static potentials above the ionosphere which accelerate the charged particles (mostly electrons) into the upper atmosphere. Because the ME at Jupiter are even brighter and \add{more} permanent, it was thought that such quasi-static potentials would also dominate the energization of the charged particles. Such quasi-static potentials have indeed seldom been discovered by Juno, but even in these specific locations, the precipitating energy flux remains dominated by stochastic processes and most electron distributions are bi-directional along the field lines \citep{mauk_diverse_2018}. This finding is a surprise since corotation enforcement models rely on the formation of such electric potentials through the Knight relationship (or a variation thereof) between the precipitating energy flux and the electron energy \citep[e.g.][]{cowley_origin_2001,ray_magnetosphere-ionosphere_2010, tao_variation_2016}. Since bi-directional electron acceleration appear\add{s} to be the norm, the UV auroral brightness, which is almost solely related to the precipitating electron energy flux, is not a reliable proxy for the intensity of the net up-going field\add{-}aligned currents. An even more unexpected and important finding is the discovery of bi-directional electron beams and proton inverted-V structures on the same field line \citep{mauk_diverse_2018, mauk_energetic_2020}, meaning that a downward current is compatible with downward moving electrons producing UV aurora. This indicates that several processes can co-exist at different altitudes on the same field line. Thus the presence of UV aurora is not even \change{an}{a sure} indication of up-going currents.

\section{Conclusions}
\label{Conclusion}
It is not expected \change{for}{that} 1-D (quasi-)stationary models \change{to}{can} explain\remove{s} all the details of the auroras at Jupiter, as they are simplifications built to better understand the most important processes at play. Nevertheless, specific predictions can be made \change{out of}{from} these models and a number of them \change{were}{are} challenged by \remove{the} measurements. \add{While adopted by most models, hypotheses such as symmetry or steady-state are not strictly required by the corotation enforcement theory. However, these assumptions quickly show their limits in the case of Jupiter and models attempting to account for local time variations} \citep{ray_local_2014, vogt_magnetotail_2020} \add{or time variability} \citep{cowley_modulation_2007} \add{provided predictions opposite to observations.} It is noteworthy that most of the observations mentioned \remove{here} above concern the generation of the auroras, which are associated \add{with}, but not strictly equivalent \add{to}, \remove{to} the magnetosphere-ionosphere coupling processes and their related currents. After all, the magnetospheric plasma at Jupiter is rotating and field\add{-}aligned current\add{s} have indeed been observed. Thus, it is not clear yet whether the question of the origin of the main emissions at Jupiter requires some adjustments of the \change{mainstream}{widely-accepted} theory \add{, the addition of other equally important mechanisms to the global explanation} or a complete paradigm shift. However, recent works have suggested possible paths forward. 
First, the idea that the auroral emissions are a direct image of the up-going field\add{-}aligned currents is invalidated by Juno's measurements \citep{mauk_diverse_2018}. If the particle acceleration process is stochastic, even regions of down-going currents would have a significant flux of down-going electrons creating auroral emissions. Then, it appears that the explanatory power of \add{(locally)} axisymmetric models is limited at Jupiter, as local time effects, fragmentation phenomena and non-axisymmetric current systems are critically important. Finally, the findings reported \remove{here} above also suggest that wave processes and wave-particle interactions should be assessed more carefully rather than assuming steady state continuous currents. A closer examination of the energy transferred by Alfv\'en waves already showed some promising results, both theoretically \citep{saur_wave-particle_2018} and observationally \citep{gershman_alfvenic_2019}. \add{Moreover, other plasma waves (auroral hiss) have also been suggested as being important in the acceleration of energetic particles in Jupiter's polar region and may also be relrevant for the ME }\citep[e.g][]{elliott_acceleration_2018,kurth_whistler_2018}. Finally, it could also be of critical importance for magneto-hydrodynamic simulations of the Jovian magnetosphere to focus on the Poynting flux and the contribution of Alfv\'en wave power rather than on the field\add{-}aligned currents when comparing their outputs to auroral images in order to provide crucial insight in understanding the origin of the main emissions. 

%%

%  Numbered lines in equations:
%  To add line numbers to lines in equations,
%  \begin{linenomath*}
%  \begin{equation}
%  \end{equation}
%  \end{linenomath*}

%% Enter Figures and Tables near as possible to where they are first mentioned:
%
% DO NOT USE \psfrag or \subfigure commands.
%
% Figure captions go below the figure.
% Table titles go above tables;  other caption information
%  should be placed in last line of the table, using
% \multicolumn2l{$^a$ This is a table note.}
%
%----------------
% EXAMPLE FIGURES
%

\acknowledgments
The authors are grateful to Pr. Jean-Claude G\'erard for helpful discussions and advise. B.B. is a Research Associate of the Fonds de la Recherche Scientifique - FNRS. B.B. and D.G. acknowledge financial support from the Belgian Federal Science Policy Office (BELSPO) via the PRODEX Programme of ESA. The data included herein are archived in NASA's Planetary Data System (\url{http://pds-atmospheres.nmsu.edu/data_and_services/atmospheres_data/JUNO/juno.html}).

%% ------------------------------------------------------------------------ %%
%% References and Citations

%%%%%%%%%%%%%%%%%%%%%%%%%%%%%%%%%%%%%%%%%%%%%%%
%
% \bibliography{<name of your .bib file>} don't specify the file extension
%
% don't specify bibliographystyle
%%%%%%%%%%%%%%%%%%%%%%%%%%%%%%%%%%%%%%%%%%%%%%%

%\bibliography{ references_zot.bib }

%Reference citation instructions and examples:
%
% Please use ONLY \cite and \citeA for reference citations.
% \cite for parenthetical references
% ...as shown in recent studies (Simpson et al., 2019)
% \citeA for in-text citations
% ...Simpson et al. (2019) have shown...
%
%
%...as shown by \citeA{jskilby}.
%...as shown by \citeA{lewin76}, \citeA{carson86}, \citeA{bartoldy02}, and \citeA{rinaldi03}.
%...has been shown \cite{jskilbye}.
%...has been shown \cite{lewin76,carson86,bartoldy02,rinaldi03}.
%... \cite <i.e.>[]{lewin76,carson86,bartoldy02,rinaldi03}.
%...has been shown by \cite <e.g.,>[and others]{lewin76}.
%
% apacite uses < > for prenotes and [ ] for postnotes
% DO NOT use other cite commands (e.g., \citet, \citep, \citeyear, \nocite, \citealp, etc.).
%

\end{document}